\documentclass[12pt,fleqn]{article}

\usepackage{amsmath}
\usepackage{amssymb}
\usepackage{epsfig}
\usepackage{graphicx,color}
\usepackage{cite}
\graphicspath{{./}}

\newcommand{\capdef}{}
\newcommand{\mycaption}[2][\capdef]{\renewcommand{\capdef}{#2}%
        \caption[#1]{{\footnotesize #2}}}
\makeatletter
\renewcommand{\fnum@table}{\textbf{\tablename~\thetable}}
\renewcommand{\fnum@figure}{\textbf{\figurename~\thefigure}}
\makeatother

\newcounter{myenumi}

\renewcommand{\themyenumi}{\roman{myenumi}}
{\end{list}}

\newlength{\myem}
\newcounter{mysubequation}[equation]

\makeatletter
\renewcommand{\section}{\@startsection{section}{1}{0em}{-\baselineskip}%
{\baselineskip}{\normalfont\large\bfseries}}
\renewcommand{\subsection}%
{\@startsection{subsection}{2}{0em}{-0.7\baselineskip}%
{0.7\baselineskip}{\normalfont\bfseries}}
\makeatother

\newcommand{\bi}{\begin{itemize}}
\newcommand{\ei}{\end{itemize}}

\newcommand{\be}{\begin{equation}}
\newcommand{\ee}{\end{equation}}
\newcommand{\bea}{\begin{eqnarray}}
\newcommand{\eea}{\end{eqnarray}}

\newcommand{\ldm}{\Delta m_{31}^2}

\newcommand{\deltacp}{\delta_{\mathrm{CP}}}
\newcommand{\stheta}{\ensuremath{\sin^2 2 \theta_{13}}}

\newcommand{\ie}{{\it i.e.}}

\newcommand{\eg}{{\it e.g.}}

\newcommand{\cf}{{\it cf.}}

\newcommand{\eq}{Eq.}
\newcommand{\eqs}{Eqs.}

\newcommand{\fig}{Fig.}
\newcommand{\Fig}{Fig.}

\newcommand{\Ref}{Ref.}
\newcommand{\Refs}{Refs.}
\newcommand{\Sec}{Sec.}

\newcommand{\Tab}{Table}

\newcommand{\figu}[1]{\fig~\ref{fig:#1}}
\newcommand{\Figu}[1]{\Fig~\ref{fig:#1}}

\begin{document}

\begin{titlepage}

\renewcommand{\thefootnote}{\alph{footnote}}

\vspace*{-1.cm}
\begin{flushright}
TUM-HEP-618/06\\
MADPH-06-1252 
\end{flushright}

\vspace*{0.5cm}

\renewcommand{\thefootnote}{\fnsymbol{footnote}}
\setcounter{footnote}{-1}

{\begin{center}
{\Large\bf From Double Chooz to Triple Chooz --- 
Neutrino Physics at the Chooz Reactor Complex} 
\end{center}}
\renewcommand{\thefootnote}{\alph{footnote}}

{\begin{center} {\large{\sc
                P.~Huber\footnote[1]{\makebox[1.cm]{Email:}
                phuber@physics.wisc.edu},~
                J.~Kopp\footnote[2]{\makebox[1.cm]{Email:}
                jkopp@ph.tum.de},~
                M.~Lindner\footnote[3]{\makebox[1.cm]{Email:}
                lindner@ph.tum.de},~
                M.~Rolinec\footnote[4]{\makebox[1.cm]{Email:}
                rolinec@ph.tum.de},~
                W.~Winter\footnote[5]{\makebox[1.cm]{Email:}
                winter@ias.edu}
                }}
\end{center}}
\vspace*{0cm}
{\it
\begin{center}

\footnotemark[1]
       Department of Physics, University of Wisconsin, \\
       1150 University Avenue, Madison, WI 53706, USA

\vspace*{1mm}

\footnotemark[2]${}^,$\footnotemark[3]${}^,$\footnotemark[4]%
       Physik--Department, Technische Universit\"at M\"unchen, \\
       James--Franck--Strasse, 85748 Garching, Germany

\vspace*{1mm}

\footnotemark[5]%
       School of Natural Sciences, Institute for Advanced Study, \\
       Einstein Drive, Princeton, NJ 08540, USA

\vspace*{1cm}

\today
\end{center}}

\vspace*{1cm}

{\Large \bf
\begin{center} Abstract \end{center}  }

We discuss the potential of the proposed Double Chooz reactor 
experiment to measure the neutrino mixing angle 
$\sin^22\theta_{13}$. We especially consider systematical 
uncertainties and their partial cancellation in a near and far 
detector operation, and we discuss implications of a delayed 
near detector startup. Furthermore, we introduce Triple Chooz, 
which is a possible upgrade scenario assuming a second, larger 
far detector, which could  start data taking in an existing 
cavern five years after the first far detector. We review the 
role of the Chooz reactor experiments in the global
context of future neutrino beam experiments. We find that 
both Double Chooz and Triple Chooz can play a leading role 
in the search for a finite value of $\sin^22\theta_{13}$. 
Double Chooz could achieve a sensitivity limit of 
$\sim 2 \cdot 10^{-2}$ at the 90\%~confidence level after 5~years while the 
Triple Chooz setup could give a sensitivity below $10^{-2}$.   

\vspace*{.5cm}

\end{titlepage}

\newpage

\renewcommand{\thefootnote}{\arabic{footnote}}
\setcounter{footnote}{0}

\section{Introduction}

Neutrino oscillations have now clearly been established for solar,
atmospheric and reactor neutrinos, as well as with neutrino
beams. However, these oscillations can still be described by an effective two 
neutrino picture to a very good approximation. This
is a consequence of the smallness of the third mixing angle
$\theta_{13}$ and the fact that the solar mass splitting $\Delta m^2_{21}$
is much smaller than the atmospheric mass splitting $\Delta m^2_{31}$.
Establishing generic three flavour effects by measuring a finite 
value for the third mixing angle $\theta_{13}$ is therefore one
of the most important tasks for future neutrino experiments. For a
concise review and description of the current status see \Ref~\cite{APS}. 
A finite value of $\theta_{13}$ is crucial for the 
search for leptonic CP violation, too. Since CP violating effects are 
proportional to $\theta_{13}$, discovering a 
finite value of $\theta_{13}$ or excluding a certain range of values 
is a key information for the planning of future long baseline neutrino 
beam experiments.
Therefore, we discuss in this paper the potential to 
limit or measure $\theta_{13}$ with Double Chooz, which is currently
the most advanced reactor project. In addition, we consider
the Triple Chooz upgrade option, which could benefit from an
existing cavern where a second large far detector could be constructed.
We also discuss how a timely information on $\stheta$ will influence 
the choice of technology for the second generation neutrino beam facilities.

The outline of the paper is as follows. In \Sec~2, we present 
some general remarks on the neutrino oscillation framework and 
we discuss implications for reactor anti-neutrino disappearance 
measurements. In \Sec~3, we describe the simulated experimental 
setups of Double Chooz and a potential upgrade to Triple Chooz. 
We then discuss in \Sec~4 the systematical errors at 
Double Chooz and we present their implementation within our 
analysis. Next, in \Sec~5, we present the results of our 
simulations for the sensitivity and the precision of 
$\sin^22\theta_{13}$. Here, we provide also a detailed 
discussion of the quantitative impact of the 
systematical uncertainties. Finally, we assess the role of
Double Chooz and eventually Triple Chooz in the global context 
of $\sin^22\theta_{13}$ measurements with reactors and future 
neutrino beam experiments.

\section{Neutrino oscillation framework}

As discussed in previous
studies~\cite{Mikaelyan:1999pm,Martemyanov:2002td,Minakata:2002jv,Huber:2003pm,Lasserre:2005qw},
reactor experiments can play a crucial role for measurements of the
third small neutrino mixing angle $\theta_{13}$. An important aspect
is that such a measurement in the $\bar{\nu}_e$-disappearance channel
does not suffer from correlations with unknown parameters, such as the CP
phase $\delta_{CP}$. Correlations with the other oscillation
parameters were also found to be negligible~\cite{Huber:2003pm}. This
can easily be seen in the expansion of the full oscillation
probability in the small parameters $\sin^22\theta_{13}$ and $\alpha
\equiv \Delta m^2_{21}/\Delta m^2_{31}$ up to second order:
\begin{equation}
1-P_{\bar{e}\bar{e}} \, \simeq \, \sin^22\theta_{13} \, \sin^2\Delta_{31} \, + \,
\alpha^2 \, \Delta_{31}^2  \cos^4\theta_{13} \, \sin^22\theta_{12}~,
\label{eq:Pee}
\end{equation}
where $\Delta_{31} = \Delta m^2_{31} L / 4E$, $L$ is the baseline, and
$E$ the neutrino energy.
Matter effects can also be safely ignored for such short baselines of 
$L=1\sim2$~km. For a measurement at the first oscillation maximum and 
$\sin^22\theta_{13}>10^{-3}$ even the second term in \eq~(\ref{eq:Pee}) 
becomes negligible\footnote{Note that the numerical simulations with 
GLoBES are not based on \eq~(\ref{eq:Pee}), but on the full three-flavour 
oscillation probability.}. Unless stated differently, we use the following
input oscillation parameters (see \eg\ \Refs~\cite{Fogli:2003th,Bahcall:2004ut,Bandyopadhyay:2004da,Maltoni:2004ei}):
\begin{eqnarray} 
  \Delta m_{31}^2 \!& = \!& 2.5 \cdot 10^{-5} \ \textrm{eV}^2~; \ \ \,
  \sin^22\theta_{23} \,  =  \, 1 \label{eq:atm}
  \label{eq:Params1} \\
  \Delta m_{21}^2 \!& = \!& 8.2 \cdot 10^{-3} \ \textrm{eV}^2~; \ \ \,
  \sin^22\theta_{12} \,  =  \, 0.83 \label{eq:sol}
  \label{eq:Params2}
\end{eqnarray}
Our analysis is performed with a modified version of the GLoBES 
Software~\cite{Huber:2004ka}, which allows a proper treatment of all kinds 
of systematical errors which can occur at a reactor experiment such as 
Double Chooz. This is important since the sensitivity of a reactor 
experiment to $\sin^22\theta_{13}$ depends crucially on these systematical 
uncertainties~\cite{Huber:2003pm}. The importance of systematical errors 
becomes obvious from \eq~(\ref{eq:Pee}), since a small quantity has to be 
measured as a deviation from~1.

\section{Experimental setups}
\label{sec:ndsim}

The basic idea of the Double Chooz experiment is a near and a far
detector which are as similar as possible in order to cancel 
systematical uncertainties. The two detectors are planned
to have the same fiducial mass of $10.16$~t of liquid scintillator.
However, there are also some unavoidable differences, such as 
 the larger muon veto in the near detector. The thermal power
of the reactor is assumed to be $2\cdot4.2$~GW (two reactor cores). The
Double Chooz setup can benefit from the existing Chooz cavern at a
baseline of $L=1.05$~km from the reactor cores. This allows a faster
startup of the far detector in order to collect as much statistics as
possible at the larger baseline.  For the near detector, a new
underground cavern must be built close to the reactor cores. In this
paper, we assume 100~m for the baseline of the near 
detector~\cite{Ardellier:2004ui}. Being so close to the reactor, it
can catch up with the statistics of the far detector. As our standard scenario 
in this paper, we assume that the near
detector starts 1.5~years after the far detector. We refer to the initial phase 
without the near detector as phase~I, and to the period in which both the
near and far detectors are in operation as phase~II.  Typically this leads for the
far detector to $19 \, 333$ unoscillated events per year, corresponding to
$1.071\cdot 10^6$ events per year in the near
detector~\cite{Ardellier:2004ui}.

Besides the Double Chooz experiment, we discuss a potential Triple Chooz
upgrade after a few years by construction of a second, larger 
far detector. Another existing cavern at roughly the same baseline 
from the Chooz reactor cores can be used for this purpose. This is a very 
interesting option, since this second cavern should be available around 2010,
and one could avoid large civil engineering costs and save time. In particular,
one could essentially spend all of the money for a typical second generation reactor experiment 
on the detector. We therefore consider a 200~t liquid scintillating 
detector with costs comparable to other proposed next generation reactor 
experiments~\cite{Goodman}. The ultimate useful size of such a detector strongly
depends on the level of irreducible systematics such as the bin-to-bin
error, which will be discussed in greater detail in the following sections.

\section{Systematical errors at Double Chooz}

A reactor neutrino experiment depends on a variety of different 
systematical errors, which are the most important limiting factor 
for $\sin^22\theta_{13}$ measurements. Any deficit in the detected 
neutrino flux could be attributed either to oscillations or to a 
different reactor neutrino flux $\Phi$. The systematical flux 
uncertainty is consequently the dominant contribution which 
must be minimized. In past experiments, the flux was deduced from 
the thermal power of the reactor, which can only be measured at the
level of a few percent. However, in next generation reactor experiments 
such as Double Chooz, a dedicated identical near detector will be 
used to precisely measure the unoscillated neutrino flux close to 
the reactor core such that the uncertainty in $\Phi$ cancels out. 
In addition, the near detector eliminates, in principle, the uncertainties 
in the neutrino energy spectrum, the interaction cross sections, 
the properties of the liquid scintillator (which is assumed to be
identical in both detectors), and the spill-in/spill-out effect. The latter 
occurs if the neutrino interaction takes place inside the fiducial 
volume, but the reaction products escape the fiducial volume or 
vice-versa. However, cancellation of systematical errors for a 
simultaneous near and far detector operation works only for the 
uncertainties that are correlated between both detectors. Any 
uncorrelated systematical error between near and far detector 
must therefore be well controlled. The knowledge of the fiducial 
detector mass or the relative calibration of normalization and 
energy reconstruction are, for instance, partly uncorrelated 
uncertainties and are therefore not expected to cancel completely.
In addition, backgrounds play a special role, as some of the 
associated uncertainties are correlated (\eg , the radioactive 
impurities in the detector), while others are not. In particular, 
since the overburden of the near detector is smaller than that of the 
far detector, the flux of cosmic muons will be higher for the near 
detector site. This requires a different design for the outer veto and 
different cuts in the final data analysis, which again introduces 
additional uncorrelated systematical errors.

Another complication in the discussion of cancellation of correlated 
uncertainties in Double Chooz is the fact that the near detector 
is supposed to start operation about 1.5~years later than the far detector. 
Therefore, only those systematical errors which are correlated 
between the detectors \emph{and} which are not time-dependent can 
be fully eliminated. This applies to the errors in the cross-sections, 
the properties of the scintillator, and the spill-in/spill-out effects.
However, it only partly applies to  systematical uncertainties in the 
background. In particular, the errors in the reactor flux and spectrum 
will be uncorrelated between phase~II, where both detectors are in 
operation, and phase~I, where only the far-detector operates. The reason for 
this is the burn-up and the periodical partial replacement of fuel 
elements. The different systematical uncertainties discussed so far 
are summarized in \Tab~\ref{tab:sys} together with their magnitudes 
we assume for Double Chooz.

\begin{table}
  \centering
  \begin{tabular}{|r|l|c|c|c|}
    \hline
      &                            & \bf Correlated & \bf Time-dependent & \bf Value for DC \\ \hline\hline
    1 & Reactor flux normalization & yes        & yes      & 2.0\%  \\ \hline
    2 & Reactor spectrum           & yes        & yes      & 2.0\% per bin \\ \hline
    3 & Cross Sections             & yes        & no       &        \\
    4 & Scintillator Properties    & yes        & no       & 2.0\%  \\
    5 & Spill-in/spill-out         & yes        & no       &        \\ \hline
    6 & Fiducial mass              & no         & no       &        \\
    7 & Detector normalization     & no         & yes      & 0.6\%  \\
    8 & Analysis cuts              & no         & no       &        \\ \hline
    9 & Energy calibration         & no         & yes      & 0.5\%  \\ \hline
   10 & Backgrounds                & partly     & partly   & 1.0\%  \\ \hline
  \end{tabular}
  \mycaption{\label{tab:sys} Systematical errors in reactor neutrino experiments (see text for
           details). The second column indicates which errors are correlated between near and far
           detector while the third column classifies which effects are time-dependent. Finally, the fourth column gives specific values we assume for the Double Chooz
           experiment.}
\end{table}

For the proper implementation of all relevant correlated and 
uncorrelated systematical uncertainties, together with an appropriate
treatment of the delayed near detector start up, we modified the 
$\chi^2$-analysis of the GLoBES Software and defined a $\chi^2$-function
which incorporates all the relevant uncertainties. The numerical simulation
assumes the events to follow the Poisson distribution, but for illustrative
purposes it is sufficient to consider the Gaussian approximation which is
very good due to the large event rates in Double Chooz. The total $\chi^2$
is composed of the statistical contributions of the far detector in phase I,
$\chi^2_{F,I}$, the far detector in phase II, $\chi^2_{F,II}$, and the
near detector in phase II, $\chi^2_{N,II}$, as well as a term
$\chi^2_\mathrm{pull}$ describing the constraints on the systematics:
\begin{align}
  \chi^2 &= \chi^2_{F,I} + \chi^2_{F,II} + \chi^2_{N,II} + \chi^2_\mathrm{pull} \, ,
  \label{eq:chi2total}
\end{align}
where
\begin{align}
  \hspace{-1 cm} \chi^2_{F,I} &= \sum_i
     \frac{\left[ (1 + a_{F,\rm fid} + a_{\rm norm} + a_{\mathrm{shape}, i}) T_{F,I,i}
       \, +\, (1 + a_{F,\rm fid} + a_{\rm bckgnd}) B_{F,I,i}
        - O_{F,I,i} \right]^2}{O_{F,I,i}}      \, ,    \label{eq:chi2F1} \\
  \hspace{-1 cm} \chi^2_{F,II} &= \sum_i
    \frac{\left[ (1 + a_{F,\rm fid} + a_{\rm norm} + a_\mathrm{drift}) T_{F,II,i}
       \, +\, (1 + a_{F,\rm fid} + a_{\rm bckgnd}) B_{F,II,i}
        - O_{F,II,i} \right]^2}{O_{F,II,i}}     \, ,    \label{eq:chi2F2} \\
  \hspace{-1 cm} \chi^2_{N,II} &= \sum_i
    \frac{\left[ (1 + a_{N,\rm fid} + a_{\rm norm} + a_\mathrm{drift}) T_{N,II,i}
        + (1 + a_{N,\rm fid} + a_{\rm bckgnd}) B_{N,II,i}
        - O_{N,II,i} \right]^2}{O_{N,II,i}}     \, ,    \label{eq:chi2N2} \\
  \hspace{-1 cm} \chi^2_\mathrm{pull} &= \frac{a_{F,\rm fid}^2}{\sigma_{F,\rm fid}^2}
       \, +\, \frac{a_{N,\rm fid}^2}{\sigma_{N,\rm fid}^2}
       \, +\, \frac{a_{\rm norm}^2}{\sigma_{\rm norm}^2}
       \, +\, \frac{a_{\rm drift}^2}{\sigma_{\rm drift}^2}
       \, +\, \frac{a_{\rm bckgnd}^2}{\sigma_{\rm bckgnd}^2}
       \, +\, \sum_i \frac{a_{\mathrm{shape},i}^2}{\sigma_{\mathrm{shape},i}^2} \, .
       \label{eq:chi2pull} 
\end{align}
In these expressions, $O_{F,I,i}$ denotes the event number in the i-th bin at
the far detector in phase~I, $O_{F,II,i}$ the corresponding event number in phase~II
and $O_{N,II,i}$ the event number in the near detector during phase~II. These event
numbers are calculated with GLoBES assuming the values given in \eqs~(\ref{eq:Params1})
and~(\ref{eq:Params2}) for the oscillation parameters. The $T_{F,I,i}$, $T_{F,II,i}$
and $T_{N,II,i}$ are the corresponding theoretically expected event numbers in the
i-th bin and are calculated with a varying fit value for $\theta_{13}$. The
other oscillation parameters are kept fixed, but we have checked that
marginalizing over them within the ranges allowed by other neutrino
experiments does not change the results of the simulations.
This is in accordance with \Ref~\cite{Huber:2003pm}.

$B_{F,I,i}$, $B_{F,II,i}$ and $B_{N,II,i}$ denote the expected background rates, which
we assume to be 1\% of the corresponding signal rates. This means in particular,
that the background spectrum follows the reactor spectrum. In reality,
backgrounds will have different spectra, however, as long as these spectra
are known, this actually makes it easier to discriminate between signal and background
because the spectral distortion caused by backgrounds will be different from that
caused by neutrino oscillations. If there are unknown backgrounds, we must
introduce bin-to-bin uncorrelated errors, which will be discussed in
section~\ref{sec:ndpot}.

As systematical errors we introduce the correlated normalization uncertainty
$\sigma_{\rm norm} = 2.8\%$ (describing the quadratic sum of the reactor flux error,
the uncertainties in the cross sections and the scintillator properties, and
the spill in/spill out effect) and the fiducial mass uncertainty for near and 
far detector $\sigma_{N,\rm fid} = 0.6\%$ and
$\sigma_{F,\rm fid} = 0.6\%$.
Furthermore, to account for errors introduced by the delayed startup of
the near detector, we allow an additional bias to the flux normalization
in phase II with magnitude $\sigma _{\rm drift} = 1\%$~per year of delay. We
also introduce a shape uncertainty $\sigma_{\mathrm{shape},i} = 2\%$~per bin
in phase~I, which describes the uncertainty in the reactor spectrum.
It is completely uncorrelated between energy bins. Note that in phase~II
a possible shape uncertainty is irrelevant as it will be canceled by
the near detector. We assume a background normalization uncertainty of
$\sigma_{\rm bckgnd} = 40\%$.
Finally, we introduce a 0.5~\% energy calibration error which is implemented
as a re-binning of $T_{F,I,i}$, $T_{F,II,i}$ and $T_{N,II,i}$ before the $\chi^2$
analysis (see App.~A of  \Ref~\cite{Huber:2003pm}). It is uncorrelated
between the two detectors, but we neglect its time dependence, since we have
checked that it hardly affects the results.

\section{Physics potential}
\label{sec:ndpot}

In this section, we present the numerical results of our analysis and
we discuss the performance of Double Chooz and the Triple Chooz 
upgrade. 
\begin{figure}[t]
  \begin{center}
    \includegraphics[width=11cm]{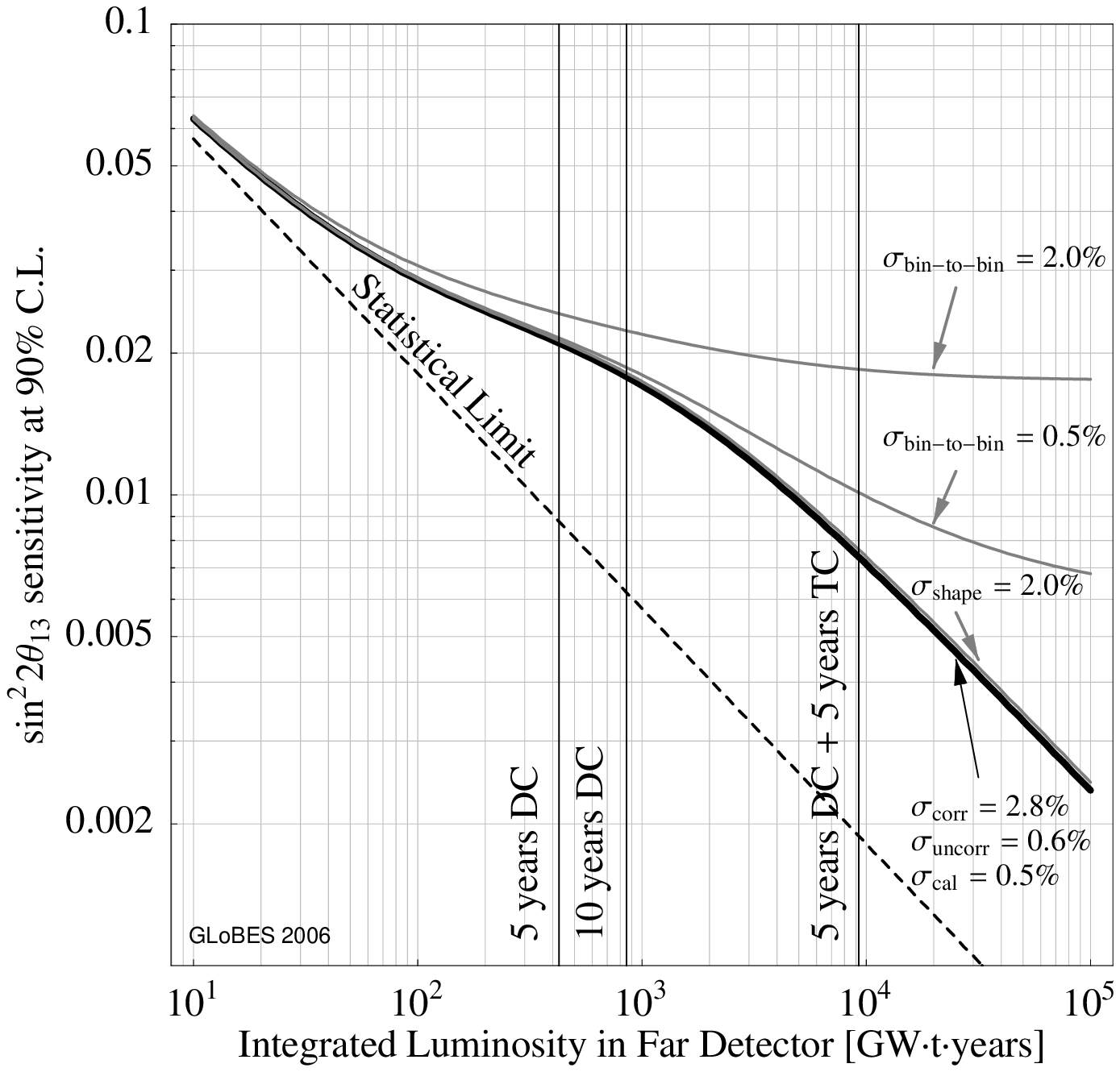} \hspace{0.5 cm}
  \end{center}
  \mycaption{\label{fig:systematics} The impact of systematical uncertainties on 
  the $\sin^22\theta_{13}$ sensitivity limit at the 90\%~confidence level as function of the 
  total integrated luminosity for a reactor experiment with near and far detector
  (both taking data from the beginning). The integrated luminosity is given by the product 
  of reactor power, far detector mass and running time in~GW~t~yrs. 
  The vertical lines indicate the exposure in 5~years of Double Chooz operation (left),
  10~years of Double Chooz (middle), and 5~years Double Chooz + 5~years Triple Chooz (right).
  We still neglect the effects of a delayed near detector startup and of the different
  baselines of the two far detectors in Triple Chooz. The plot illustrates that for high
  luminosities it is crucial to control the uncorrelated uncertainties, in particular the
  bin-to-bin errors.}
\end{figure}
First, we discuss the quantitative impact of the systematical uncertainties 
introduced in the last section.
In \figu{systematics}, we assume a reactor experiment with identical near and 
far detectors located at a baseline of 1.05~km which are running
simultaneously. Note that this is neither the initial Double Chooz setup, where
the near detector will be added with some delay, nor the Triple Chooz setup, 
which would have two different far detectors at slightly different baselines.
\Figu{systematics} is nevertheless interesting, since it allows to
compare the principal strength of the Double Chooz and Triple Chooz
setups. The vertical black lines in \figu{systematics} correspond to 
5~years of {\em full} Double Chooz operation
(5~yrs~$\times$~10.16~t~$\times$~8.4~GW),
10~years of {\em full} Double Chooz operation
(10~yrs~$\times$~10.16~t~$\times$~8.4~GW)
and 5~years of {\em full} Double Chooz + 5~years Triple Chooz 
([5~yrs~$\times$~10.16~t + 5~yrs~$\times$~210.16~t]~$\times$~8.4~GW),
respectively. 
The sensitivity of an experiment with the integrated luminosity of
$\sim 10^3 \mathrm{GW}\,\mathrm{t}\,\mathrm{yrs}$, such as Double Chooz, is
  quite independent of the bin-to-bin error as can be seen from
  \figu{systematics}. This is not surprising and has been already
  discussed in detail in \Ref~\cite{Huber:2003pm}. Therefore, a sensitivity
  down to $\stheta=0.02$ is certainly obtainable.
The situation is somewhat different for an experiment of the size of
Triple Chooz. From discussions in \Ref~\cite{Huber:2003pm} it is expected
that the $\sin^22\theta_{13}$ sensitivity limit at a reactor experiment 
of the size of Triple Chooz should be quite robust with respect to 
systematical uncertainties associated to the normalization, since 
the normalization is determined with good accuracy from the very 
good statistics and from additional spectral information. This robustness 
can be seen in \figu{systematics} where the sensitivity limit at the 
90\%~confidence level for different sets of systematical errors is shown as function
of the total integrated luminosity in the far detector (given by the product 
of reactor power, detector mass and running time in~GW~t~yrs). 
As can be seen in \figu{systematics}, the performance at luminosities
associated with Triple Chooz decreases immediately if in addition bin-to-bin
errors are introduced which are uncorrelated between near and far
detector. These uncorrelated bin-to-bin errors are added to the
$\chi^2$ function in the same way as $\sigma_{\rm shape}$ was introduced in
\eqs~(\ref{eq:chi2total}) to~(\ref{eq:chi2pull}) for each bin independently,
but uncorrelated between the two detectors. These uncertainties could, for
instance, come from uncorrelated backgrounds and different cutting
methods necessary if the detectors are not 100\%\ 
identical. Thus, especially for the Triple Chooz setup, these
uncertainties have to be under control, because they can spoil the
overall performance. The bin-to-bin error is used here as a
parameterization for yet unknown systematical effects and is an attempt
to account for the worst case. Thus, in a realistic situation, the
bin-to-bin error would have to be broken down into individual known
components and thus the impact would be less severe.  If bin-to-bin
errors were excluded, the evolution of the sensitivity limit would 
already enter a second statistics dominated regime (curve parallel
to dashed statistics only curve), since the systematical uncertainties
could be reduced due to the spectral information in the data (see also
\Ref~\cite{Huber:2003pm} for explanations).
Note that the $\sigma_{\rm shape}$ uncertainty does not affect the 
$\sin^22\theta_{13}$ sensitivity in a sizeable manner, since it is 
correlated between near and far detector and therefore cancels out. 

\begin{figure}[t]
  \begin{center}
    \includegraphics[width=11 cm]{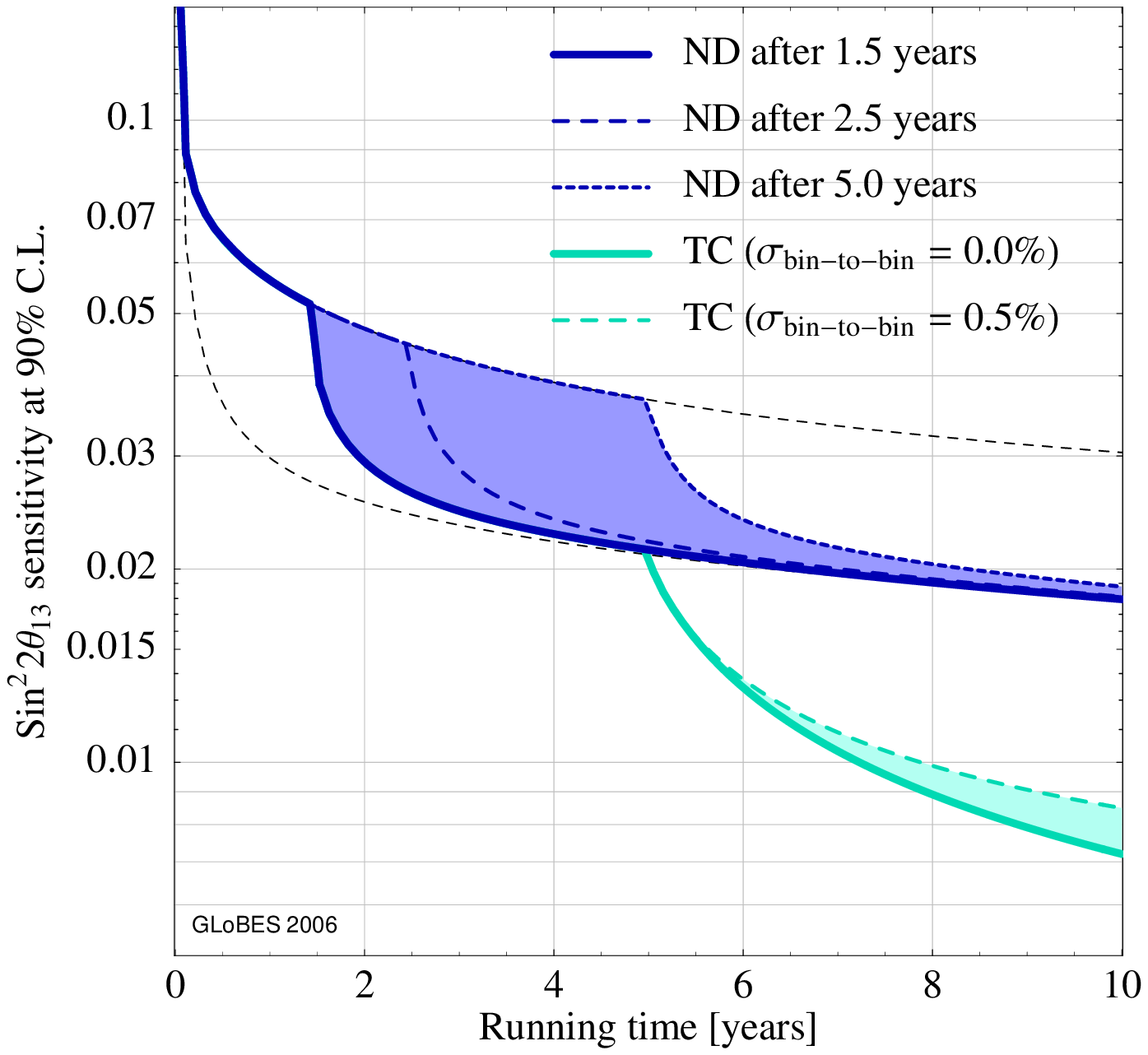} \hspace{0.5 cm}
  \end{center}
  \mycaption{\label{fig:dc-spectral} The $\sin^22\theta_{13}$ sensitivity limit at the 90 \%
  confidence level achievable at Double Chooz for three different delayed startup times of the 
  near detector, and of the Triple Chooz Scenario, where the second far detector is
  added after 5~years of Double Chooz running.}
\end{figure}
The evolution of the $\sin^22\theta_{13}$ sensitivity limit at 
the 90\% confidence level as a function of the running time is 
shown in \figu{dc-spectral}. Here the upper thin dashed curve 
indicates the limit which could be obtained by the far detector 
of Double Chooz alone (\ie, no near detector is assumed,
which corresponds to phase~I continuing up to 10~years), while the 
lower thin dashed curve shows the limit which could be obtained 
if the near detector started data taking together with the far
detector (\ie, phase~I is absent, while
phase~II continues up to 10~years). The near detector improves 
the sensitivity considerably, but even the far detector alone 
would quickly improve the existing Chooz limit. The solid blue (black) curve
corresponds to the standard Double Chooz scenario, where the near detector
starts operation 1.5~years after the far detector. It can be seen that the
$\sin^22\theta_{13}$ limit improves strongly after the startup of the near
detector and converges very fast to the curve corresponding to a near detector 
in operation from the beginning. Thus, the Double Chooz performance 
does not suffer from the delayed near detector startup in the end. 
This ``delayed startup'' is in fact not a delay, but it allows 
a considerably quicker startup of the whole experiment, utilizing
the fact that no civil engineering is necessary at the site of 
the far detector. There have been performed similar calculations by the Double 
Chooz collaboration~\cite{lasserre:private}, concerning the evolution of the 
$\stheta$ sensitivity with a 1.5 years duration of phase I, followed by a 
phase II scenario, which are in good agreement with the corresponding 
curves in \figu{dc-spectral}. However, there are slight differences especially for 
the evolution of the $\stheta$ sensitivity in phase I. These come from the
inclusion of spectral information in \figu{dc-spectral}, 
whereas in the calculations in \Ref~\cite{lasserre:private} only total rates 
were taken into account. 
The dashed and dotted blue (black) 
curves in \figu{dc-spectral} show the evolution of the sensitivity 
limit, if the near detector were operational not 1.5~years after 
the far detector, but 2.5 or 5~years, respectively. Again, the 
sensitivity limit improves quickly as soon as the near detector is 
available and quickly approaches the limit with a near detector from 
the beginning. The main reason for this is, that the overall 
sensitivity is ultimately dominated by the uncorrelated systematical 
uncertainties and not by statistics. 
Furthermore, \figu{dc-spectral} shows the evolution of the 
$\sin^22\theta_{13}$ sensitivity limit for the Triple Chooz setup,
both without uncorrelated bin-to-bin errors (solid cyan/grey curve) and with
$\sigma_\textrm{bin-to-bin} = 0.5\%$ (dashed cyan/grey curve). It is assumed
that the second far detector starts operation 5~years after the first
far detector. In the  Triple Chooz simulation, we
have assumed the uncorrelated normalization and energy calibration
errors of the second far detector to be 1\% each. This is slightly larger
than the 0.6\% resp. 0.5\% in the original Double Chooz reflecting that
the design of the new detector would have to be different from that of the 
two original detectors. It can be seen that the Triple Chooz 
scenario could achieve a 90\%~confidence level sensitivity limit below 
$\sin^22\theta_{13}=10^{-2}$ after less than 8~years of total running 
time (5~years Double Chooz + 3~years Triple Chooz), even if small
bin-to-bin errors were allowed to account for backgrounds or detector
characteristics that are not fully understood. If bin-to-bin errors
are absent, the sensitivity will improve by about 10\%. The plot shows that 
the Triple Chooz setup can compete with the sensitivity expected from other 
second generation precision reactor experiments. It also demonstrates that
the precision of reactor experiments could be further improved in a
timely manner. The improved $\sin^22\theta_{13}$ limits
or measurements could be valuable input for planning and optimizing the
second generation neutrino beam experiments.

We have so far considered a 200t detector for Triple Chooz
and one may wonder how an even larger detector, which
easily fits into the large existing cavern, would perform.
A larger detector implies even higher values of integrated
far detector luminosity. From \Figu{systematics} one can immediately see
that the achievable value of $\sigma_{\rm bin-to-bin}$ determines
the performance, \ie\ if one can benefit from the larger detector mass
or if the sensitivity is already saturated by $\sigma_{\rm bin-to-bin}$.
From \Figu{systematics} one can read off that for 100t or 200t 
$\sigma_{\rm bin-to-bin} < 0.5\%$ should be achieved. A 500t detector
would require $\sigma_{\rm bin-to-bin} < 0.1\%$ in order to obtain 
an improvement of the sensitivity limit to the level of $5\cdot 10^{-3}$.

\begin{figure}[t]
  \begin{center}
    \includegraphics[width=11 cm]{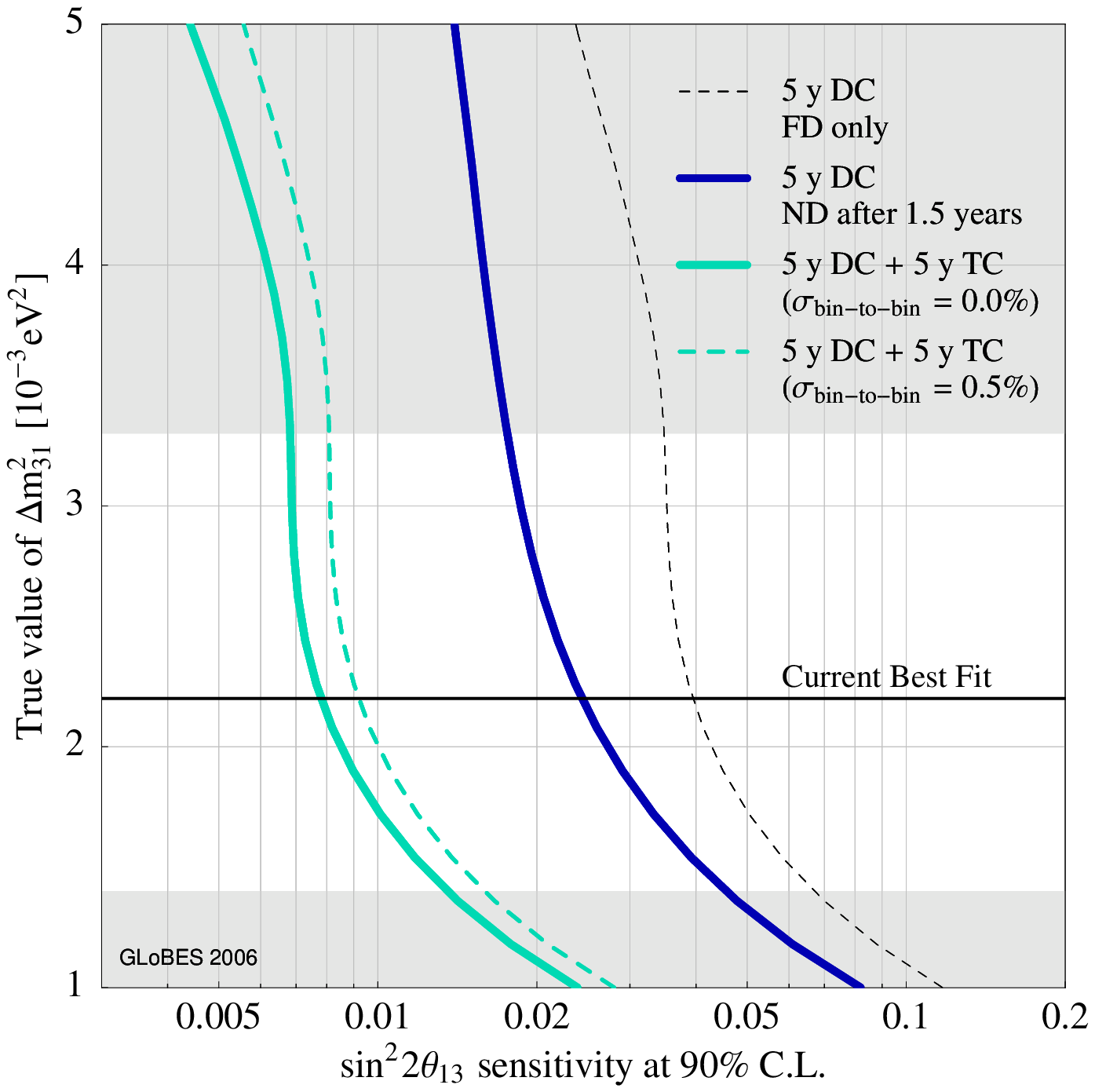} \hspace{0.5 cm}
  \end{center}
  \mycaption{\label{fig:dm2} The sensitivity limit for $\sin^22\theta_{13}$ 
             at the 90\%~confidence level as a function of the true value of $\ldm$. The curves
             correspond to the following setups: a 5-year run of only the 
             Double Chooz far detector without near detector (dashed blue/black
             curve to the right), a 5-year run of Double Chooz with near detector
             after 1.5~years (solid blue/black curve), and a 5-year run of Double
             Chooz followed by a 5-year run of Triple Chooz without bin-to-bin
             errors (solid cyan/grey curve) and with a 0.5\% bin-to-bin error
             (dashed cyan/grey curve). The light grey areas show the $3\sigma$ excluded 
             regions for $\ldm$ from a global fit \cite{Maltoni:2004ei},
             the horizontal line indicates the corresponding best fit value.}
\end{figure}
\Figu{dm2} shows the dependence of the $\sin^22\theta_{13}$ sensitivity of 
Double Chooz on the true value of $\ldm$. Such a 
parametric presentation makes sense, since $\ldm$ will be known
relatively precisely by then from the MINOS experiment. 
The sensitivity again is shown for four different scenarios: 5~years
with the far detector of Double Chooz only (dashed black curve to the right), 
5~years of Double Chooz with a near detector after 1.5~years
(solid blue/black curve), and finally the Triple Chooz scenario with
and without bin-to-bin errors, where the second far detector is 
starting operation 5~years after the first far detector (cyan/grey curves).
We also show the curves for a region of $\ldm$ parameter space that is
already excluded by current global fits (upper grey-shaded region;
see, \eg, \Refs~\cite{Fogli:2003th,Bahcall:2004ut,Bandyopadhyay:2004da,Maltoni:2004ei}).
One can easily see that a larger true value of $\ldm$ would be favorable for an
experiment at the relatively short baseline of $L \sim 1.05$~km between the reactor and the
Double Chooz detector.
As can be seen in \figu{dm2}, the setup with only a far detector and the 
Triple Chooz setup show a characteristic dip around 
$\ldm \approx 3 \cdot 10^{-3} \, \mathrm{eV}^2$. This effect is due to the normalization 
errors and can be understood as follows: If the true $\ldm$ is very small, the 
first oscillation maximum lies outside the energy range of reactor neutrinos. 
For $\ldm \approx 2 \cdot 10^{-3} \, \mathrm{eV}^2$, the first maximum enters at the 
lower end of the spectrum. Therefore oscillations cause a spectral distortion
which cannot be mimicked by an error in the flux normalization. But with 
increasing true $\ldm$, a larger part of the relevant energy range is affected 
by the oscillations. This behaviour could also come from a normalization
error which decreases the sensitivity to $\stheta$ in the region around
$\ldm \approx 3 \cdot 10^{-3} \, \mathrm{eV}^2$. For even larger 
$\ldm \gtrsim 4 \cdot 10^{-3} \, \mathrm{eV}^2$, the second oscillation maximum enters
the reactor spectrum, which again causes a characteristic spectral distortion.

Up to now, we have only considered the achievable $\sin^22\theta_{13}$
sensitivity limit. If a finite value were observed, reactor
experiments could determine $\sin^22\theta_{13}$ with a certain
precision, since no correlations with the unknown CP phase
$\delta_{CP}$ would exist. For a large reactor experiment, this might allow the first 
generation beam experiments, T2K and NOvA to have a first glimpse on CP
violation~\cite{Huber:2004ug}.  
\begin{figure}[t]
  \begin{center}
    \includegraphics[width=11cm]{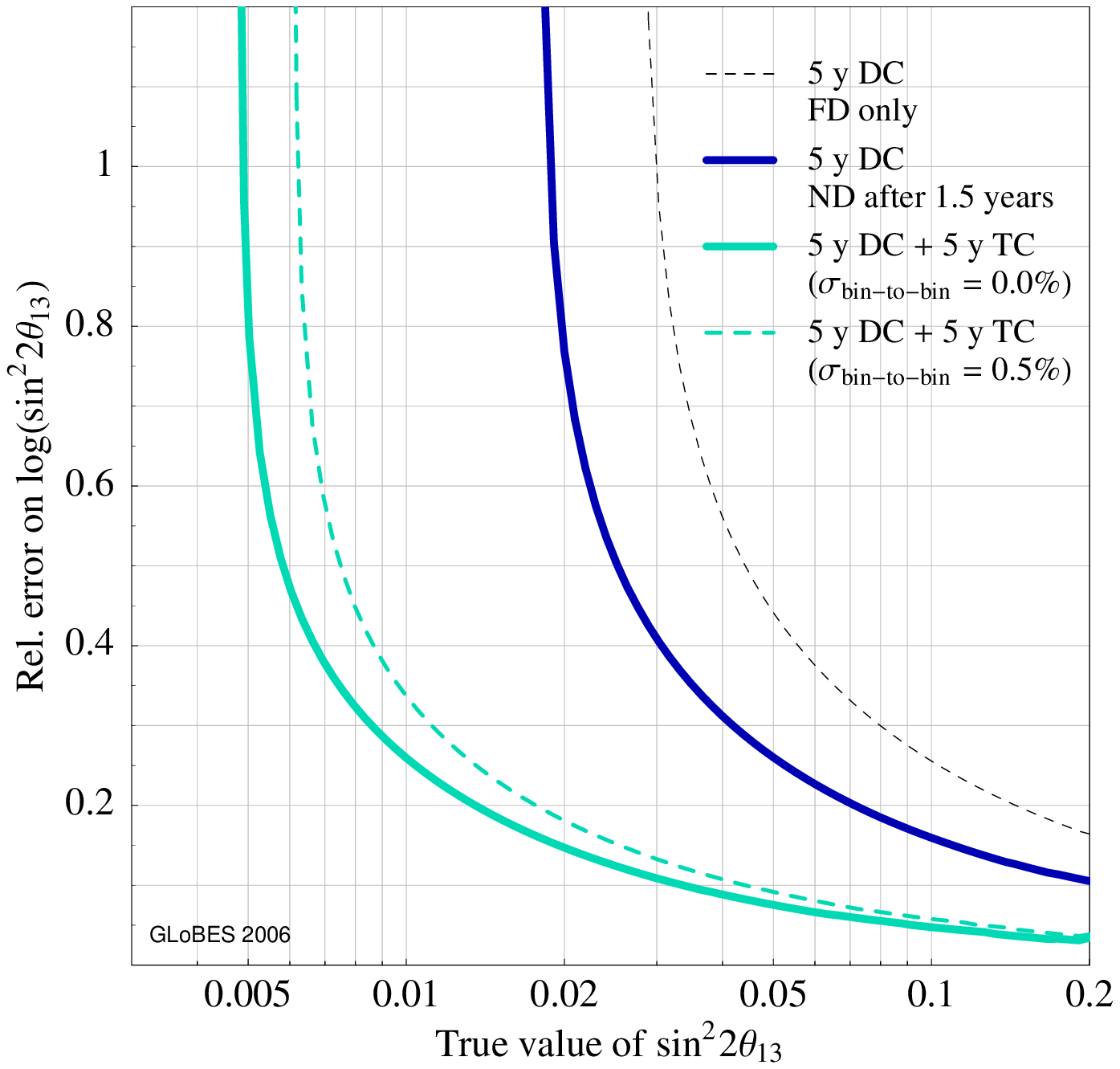} \hspace{0.5 cm}
  \end{center}
  \mycaption{\label{fig:precision} The precision of the $\sin^22\theta_{13}$ 
   measurement at the 90\%~confidence level as a function of the true value of 
   $\sin^22\theta_{13}$.  The curves correspond to the following setups: a 5-year
   run of only the Double Chooz far detector without near detector (dashed blue/black
   curve to the right), a 5-year run of Double Chooz with near detector after 1.5~years
   (solid blue/black curve), and a 5-year run of Double Chooz followed by a 5-year
   run of Triple Chooz without bin-to-bin errors (solid cyan/grey curve) and with
   a 0.5\% bin-to-bin error (dashed cyan/grey curve).}
\end{figure}
\Figu{precision} shows the precision to $\stheta$
for the different considered setups. This precision is
defined as
\begin{equation}
  \textrm{Rel. error on \stheta} = \left| \frac{\log(\sin^2 2\theta_{13}^{(u)})
             - \log(\sin^2 2\theta_{13}^{(d)})}{\log(\sin^2 2\theta_{13}^{(\rm true)}}) \right|~,
\end{equation}
where $\log(\sin^2 2\theta_{13}^{(u)})$ and $\log(\sin^2 2\theta_{13}^{(d)})$ 
are the upper and lower bounds of the 90\% confidence region, and 
$\log(\sin^2 2\theta_{13}^{(\rm true)})$ is the true value assumed in 
the simulation (same definition as in \Ref~\cite{Huber:2003pm}). The plot 
confirms the expectation that the precision is better for a larger value of $\stheta$. 
The ability to measure 
$\sin^22\theta_{13}$ is then completely lost for true  values near the 
sensitivity limit.

\section{Role in the global context and complementarity to beam experiments}

\begin{figure}[t!]
\begin{center}
\includegraphics[width=\textwidth]{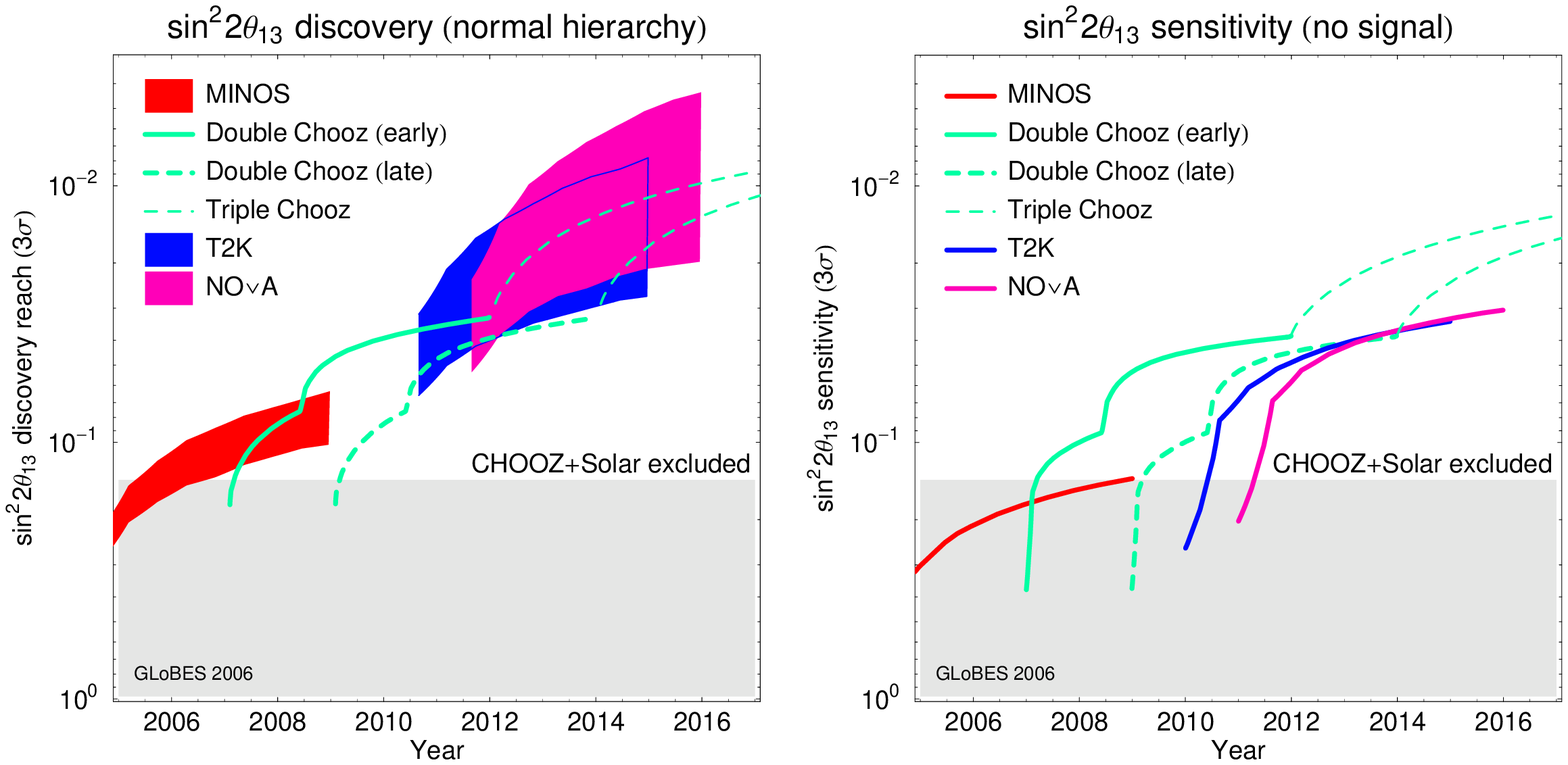}
\end{center}
\mycaption{\label{fig:evolution} A possible evolution of the
   $\stheta$ discovery potential (left) and $\stheta$ sensitivity/exclusion
   limit (right) at 3$\sigma$ as function of time including statistics, 
   systematics, and correlations ($3 \sigma$). The bands reflect 
   for the neutrino beam experiments the dependence on the unknown 
   value of $\deltacp$, \ie, the actual sensitivity evolution will lie
   in between the best case (upper) and worst (lower) curve depending 
   on the value of $\deltacp$ chosen by nature. All experiments are 
   assumed to be operated five years and the beam experiments are
   operated with neutrino running only. The full detector mass is 
   assumed to be available right from the beginning for the beam 
   experiments, \ie, the starting times are chosen accordingly.
   Double Chooz is assumed to start data taking with the near 
   detector 1.5~years after the far detector, where two possible far 
   detector starting times are shown. In addition, the possible upgrade
   to Triple Chooz is included after five years of data taking.
   Though the starting times of the experiments have been chosen as 
   close as possible to those stated in the respective LOIs, they have 
   to be interpreted with care. A normal mass hierarchy is assumed for 
   this plot and for an inverted hierarchy, the accelerator-based
   sensitivities are expected to shift down somewhat. The calculations 
   (including time evolution) of the beams are based on the experiment 
   simulations in
   \Refs~\cite{Huber:2002mx,Huber:2002rs,Huber:2004ug,Huber:2004ka} 
   using GLoBES~\cite{Huber:2004ka}.
   This Figure is similar to the ones that can be found in \Refs~\cite{Albrow:2005kw,PDNOD}.}
\end{figure}

In order to discuss the role of the Double Chooz and Triple Chooz setups in the
global context, we show in \figu{evolution} a possible evolution of the $\stheta$
discovery potential (left) and $\stheta$ sensitivity limit (right) as
function of time. In the left panel of \figu{evolution}, we assume
that $\stheta$ is finite and that a certain unknown value of
$\deltacp$ exists.  The bands in the figure reflect the dependence on
the unknown value of $\deltacp$, \ie, the actual sensitivity will lie
in between the best case (upper) and worst (lower) curve, depending on
the value of $\deltacp$ chosen by nature. In addition, the curves for
the beam experiments shift somewhat to the worse for the inverted mass
hierarchy, which, however, does not qualitatively affect this
discussion.
The right panel of the figure shows the $\stheta$ limit which can be
obtained for the hypothesis $\stheta=0$, \ie, no signal. Since 
particular parameter combinations can easily mimic $\stheta=0$ in 
the case of the neutrino beams, their final $\stheta$ sensitivity 
limit is spoilt by correlations (especially with $\deltacp$) compared 
to Double Chooz\footnote{Note that we define the $\stheta$ sensitivity limit 
as the largest value of $\stheta$ which fits (the true) $\stheta=0$
at the given confidence level. Therefore, this definition has no 
dependence on the true value of $\deltacp$, and the fit $\deltacp$ 
is marginalized over (\cf, App.~C of \Ref~\cite{Huber:2004ug}).}.
The two panels of \figu{evolution} very nicely illustrate the 
complementarity of beam and reactor experiments: Beams are sensitive 
to $\deltacp$ (and the mass hierarchy for long enough baselines), 
reactor experiments are not. On the other hand, reactor experiments 
allow for a ``clean'' measurement of $\stheta$ without being affected 
by correlations.

There are a number of important observations which can be read off
from \figu{evolution}. First of all, assume that Double Chooz starts
as planned (solid Double Chooz curves). Then it will quickly exceed
the $\stheta$ discovery reach of MINOS, especially after the near
detector is online (left panel). For some time, it would certainly be
the experiment with the best $\stheta$ discovery potential. If a
finite value of $\sin^22\theta_{13}$ were established at Double Chooz,
the first generation superbeam experiments T2K and NOvA could try to
optimize a potential anti-neutrino running strategy. The breaking of parameter 
correlations and degeneracies might in this case be even achieved by the synergy 
with the Triple Chooz upgrade (similar to Reactor-II in \Ref~\cite{Huber:2003pm}).  
For the $\stheta$ sensitivity, \ie, if there
is no $\stheta$ signal, the best limit will come from Double Chooz
already from the very beginning even without near detector. Together
with the near detector, this sensitivity cannot be exceeded by the
superbeams without upgrades, because these suffer from the correlation
with $\deltacp$.  Double Chooz has altogether an excellent chance to
observe a finite value of $\theta_{13}$ first. If $\theta_{13}$ were
zero or tiny, then Double Chooz would be an extremely good exclusion
machine. It could exclude a large fraction of the parameter space
already a few years before the corresponding superbeams.

One can also read off from \figu{evolution} that the starting time of
the near detector of Double Chooz is time-critical (\cf, dashed Double
Chooz curves). Especially from the left panel, one can see that the
near detector has to start taking data considerably before 2010 in
order to be competitive to the superbeams. Also, to achieve the
maximal synergy, it will be important that the information from Double
Chooz is available roughly around 2010 as we will discuss later on.
Note that the superbeams do not in all cases have a better $\stheta$
discovery potential. This holds especially if the true $\deltacp \sim
\pi/2$ (\cf, \eg, \Ref~\cite{Winter:2003ye}) and the mass hierarchy is
inverted. In this case, Double Chooz may still discover $\stheta$ if the near
detector starts 2010 or later. However, this scenario only holds for a
very small fraction of the parameter space.

\begin{figure}[t!]
  \begin{center}
    \includegraphics[width=11cm]{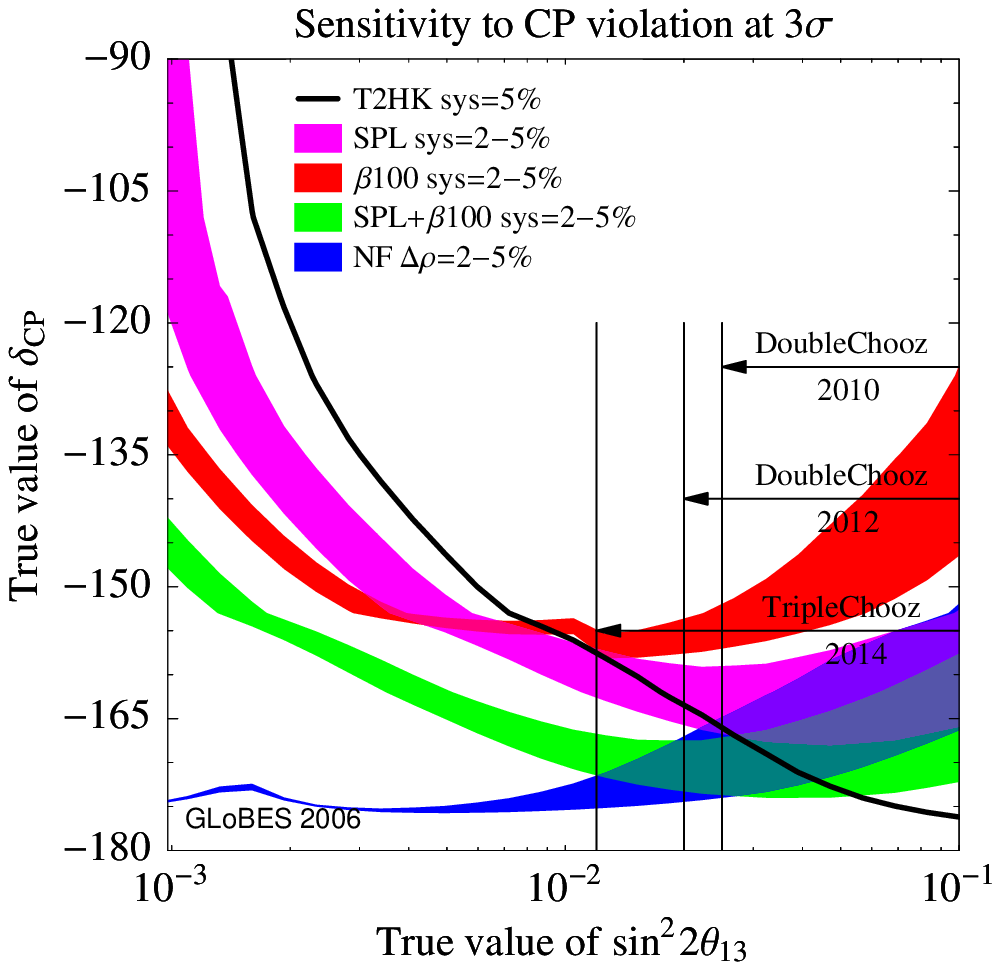} \hspace{0.5 cm}
  \end{center}
  \mycaption{\label{fig:beams} $\sin^22\theta_{13}$ sensitivity limits
    at the $90\%$~confidence level of Double and Triple Chooz in comparison to the
    $3\,\sigma$ discovery reaches (above curves) for CP violation of various, second
    generation beam experiments. 
    All curves have been calculated with
GLoBES~\cite{Huber:2004ka} including correlations and degeneracies. For all setups, 
the appropriate disappearance
channels have been included. The beta beam is
    lacking muon neutrino disappearance, which is replaced by a 
    10\% precision on $\Delta m_{31}^2$ (corresponding to the 
    T2K disappearance information).
In all cases systematics
between neutrinos, anti-neutrinos, appearance, and disappearance is
uncorrelated. For all setups with a water Cherenkov detector, the
systematics applies both to background and signal, uncorrelated.  
The neutrino factory (NF) assumes $3.1\cdot10^{20}$ $\mu^+$ decays per year for
10 years and $3.1\cdot10^{20}$ $\mu^-$ decays for 10 years. It has one
detector with $m=100\,\mathrm{kt}$ at $3000\,\mathrm{km}$ and another
detector with $30\,\mathrm{kt}$ at $7000\,\mathrm{km}$. 
The density errors between the two baselines are uncorrelated. The systematics are
0.1\% on the signal and 20\% on the background, uncorrelated. 
The detector threshold and the other parameters are taken
from \Ref~\cite{Huber:2002mx} and approximate the results
of \Ref~\cite{Cervera:2000vy}.  The beta beam ($\beta 100$) assumes $5.8\cdot10^{18}$ He
decays per year for five years and $2.2\cdot 10^{18}$ Ne decays per
year for five years. The detector mass is $500\,\mathrm{kt}$. The
detector description and the glb-file is from \Ref~\cite{Campagne:2005jh}.
The SPL setup is taken from \Ref~\cite{Mezzetto:2005ae}, and the detector
mass is $500\,\mathrm{kt}$.  The T2HK setup is taken
from \Ref~\cite{Huber:2002mx} and closely follows the
LOI~\cite{Itow:2001ee}. The detector mass is $1 \, 000\,\mathrm{kt}$ and
it runs with $4\,\mathrm{MW}$ beam power, 6 years with anti-neutrinos
and 2 years with neutrinos. The systematic error on both background
and signal is 5\%.
}
\end{figure}

Triple Chooz is a very interesting upgrade option for Double Chooz. 
\Figu{evolution} shows that it could play an important role, since 
it would have a sensitivity reaching into the discovery range of 
the neutrino beam experiments T2K and NO$\nu$A. In the case
of a value of $\stheta$ not too far below the current CHOOZ bound,
this might even lead to the possibility to restrict the CP
parameter space. Note, however, that a delayed startup would eliminate
the $\stheta$ discovery opportunity. In either case, if the true $\theta_{13}$ 
is small, Double Chooz  with a later Triple Chooz upgrade will give the best 
exclusion limits for the coming 10~years.
Note that in the staged approach of Triple Chooz, the original Double
Chooz experiment serves as a testbed for the upgrade. Thus, systematical
uncertainties will be well understood, so that a reliable sensitivity
prediction for Triple Chooz will be possible.

Double Chooz and Triple Chooz will play a central role in  selecting the optimal technology
for the second generation beam experiments. \figu{beams} shows the sensitivity to
CP violation at $3\,\sigma$ confidence level  ($\Delta\chi^2=9$) for several approaches
that are currently being discussed. Sensitivity to CP violation is
defined, for a given point in the $\theta_{13}$-$\delta$-plane (above curves), by
being able to exclude $\delta=0$ and $\delta=\pi$ at the given
confidence level. 
In \figu{beams}, clearly two regimes can be distinguished:
very large $\sin^22\theta_{13}\geq0.01$ and very small
$\sin^22\theta_{13}\leq0.01$. At large $\theta_{13}$, the sensitivity
to CP violation is basically completely determined by factors such as
systematic errors or matter density uncertainty. Thus the question of
the optimal technology cannot be answered with confidence at the
moment, since for most of the controlling factors the exact magnitude
can only be estimated. The technology decision for large
$\theta_{13}$, therefore, requires considerable R\&D. On the other
hand, in the case of small $\theta_{13}$ the optimal technology seems
to be a neutrino factory\footnote{Not shown in figure~\ref{fig:beams}
  is the $\gamma=350$ beta beam~\cite{Burguet-Castell:2005pa}, which could play the
  role of a neutrino factory.} quite independently from any of the
above mentioned factors. The branching point between the two regimes
is around $\sin^22\theta_{13}\sim0.01$ which coincides with the
sensitivities obtainable at the Chooz reactor complex. Moreover, the
information from Chooz would be available around 2010 which is
precisely the envisaged time frame for the submission of a proposal
for those second generation neutrino beam facilities. Thus the Double
Chooz results are of central importance for the long term strategy of
beam-based neutrino physics.

\section{Summary and conclusions}

We have analyzed the physics potential of new reactor
neutrino experiments at the Chooz reactor complex. A first very
realistic and competitive option is the Double Chooz project. Our
simulations show that it could be the leading experiment in the search
for a finite value of $\stheta$ in the coming years.  Therefore, Double
Chooz, if timely performed, has excellent chances to detect the first signal of a finite
value of $\stheta$. Such an early discovery would be very important
for the superbeam optimization in terms of antineutrino running to discover
mass hierarchy and CP violation, and the choice of the optimal technology
for second-generation superbeams or beta beams.  
Provided that $\stheta \gtrsim 0.04$, Double Chooz would
provide such an early signal for a finite value of $\stheta$ at relatively low cost.
In addition, Double Chooz can provide and dominate an excellent
limit for $\stheta$ if $\stheta$ is very small, because it is hardly
affected by correlations.  In the case of an exactly vanishing true
value of $\stheta=0$, Double Chooz could set an upper limit of
$\stheta<0.018$ at the 90\% confidence level after 10 years which can hardly
be exceeded by the superbeams.  However,
Double Chooz will not replace the need for the superbeams, because
superbeams have a much better $\stheta$ discovery potential and, if
$\stheta$ is large, are sensitive to $\deltacp$ and the neutrino mass
hierarchy. In summary, Double Chooz is an exclusion machine,
whereas superbeams are discovery machines, both providing very
complementary information.

We have also discussed a very interesting upgrade option for Double Chooz,
which we call ``Triple Chooz''. Similar to Double Chooz, Triple Chooz could
benefit from an existing underground cavern, which would reduce the
costs significantly. The existing cavern would also allow a faster realization of the
experiment, since no major civil construction would be necessary.
Triple Chooz would also benefit from the existing experience and
infrastructure of Double Chooz.  Our simulations show that the Triple
Chooz upgrade could compete with other planned second generation
reactor experiments.  Triple Chooz with a fiducial mass of $200 \, \mathrm{t}$
could measure $\stheta$ with a
precision better than 10\% at the 90\% confidence level down to true
values of $\stheta \simeq 0.04$ and achieve a sensitivity level well
below $\stheta \lesssim 10^{-2}$ if the true value is zero.  
Arriving early at this ``branching point'' could be very important for the technology
choice between a superbeam upgrade and neutrino factory (or higher gamma beta beam) program.
In summary, our study
shows that both Double Chooz and Triple Chooz would be very well
positioned in the global neutrino oscillation program.

\subsection*{Acknowledgments}

We would like to thank H. de Kerret, Th. Lasserre, G. Mention
and A. Merle for discussions and useful comments. This work has been
supported by SFB 375 and the Graduiertenkolleg 1054 of Deutsche
Forschungsgemeinschaft. JK would like to acknowledge support from
the Studienstiftung des deutschen Volkes. WW would like to acknowledge
support from the W.~M.~Keck Foundation and NSF grant PHY-0503584. PH
would like to acknowledge the warm hospitality of the TUM.

\newpage

\bibliographystyle{./apsrev} \bibliography{./references}

\end{document}